\documentclass[12pt]{article}
\usepackage{amssymb,amsmath,epsfig,cite}
\allowdisplaybreaks 
\allowdisplaybreaks 
\begin{document}
\title{\textbf{Charged Gravastars in Modified Gravity}}
\author{Z. Yousaf$^1$\thanks{zeeshan.math@pu.edu.pk}, Kazuharu Bamba$^2$
\thanks{bamba@sss.fukushima-u.ac.jp}, M. Z. Bhatti$^1$\thanks{mzaeem.math@pu.edu.pk} and U. Ghafoor$^1$\thanks{usamaghafoor62@gmail.com}\\
$^{1}$ Department of Mathematics, University of the Punjab,\\
Quaid-i-Azam Campus, Lahore-54590, Pakistan\\
$^2$ Division of Human Support System,\\ Faculty of Symbiotic Systems Science,\\
Fukushima University, Fukushima 960-1296, Japan}
\date{}
\maketitle
\begin{abstract}
In this paper, we investigate the effects of electromagnetic field
on the isotropic spherical gravastar models in metric $f(R,T)$
gravity. For this purpose, we have explored singularity-free
exact models of relativistic spheres with a specific equation of state.
After considering Reissner Nordstr\"{o}m spacetime as an exterior region, the interior charged manifold is matched
at the junction interface. Several viable realistic characteristics of the spherical gravastar model are studied in the presence of
electromagnetic field through graphical representations. It is concluded that the electric charge has
a substantial role in the modeling of proper length, energy contents, entropy and
equation of state parameter of the stellar system. We have also explored the stable regions of the charged gravastar structures.
\end{abstract}
{\bf  numbers:} 04.70.Dy; 04.70.Bw; 11.25.-w

\section{Introduction}

Recent interesting outcomes from a few observational
experiments like, cosmic microwave background (CMB)
radiation, Supernovae type Ia, etc. \cite{ya1} claimed that our cosmos is in the phase of accelerating expansion. The
observational ingredients from the BICEP2
experiment~\cite{Ade:2014xna, Ade:2015tva, Array:2015xqh}, the Planck
satellite~\cite{ya2, Planck:2015xua, Ade:2015lrj}, and the
Wilkinson Microwave anisotropy probe (WMAP)~\cite{Komatsu:2010fb,
Hinshaw:2012aka}, state that our cosmos is composed of only 5\% baryonic matter, the rest
is comprising of dark matter (DM) and dark energy (DE). Their percentages are observed to be 27\% and 68\%,
respectively.

A popular approach to understand the structure formation and evolution of the Universe
is modified gravity theories. Such theories are obtained by generalizing the
usual Einstein-Hilbert action. The importance as well as the need of such theories have been
discussed in detail by Nojiri and Odintsov \cite{ya3}. The $f(R)$ ($R$ is the Ricci scalar) \cite{fR1}, $f(\mathcal{T})$ ($\mathcal{T}$ is the torsion scalar) \cite{fT1}, $f(R,\Box R, T)$ ($\Box$ is the de Alembert's operator and $T$ is the trace of energy momentum tensor) \cite{box1}, $f(G)$ ($G$ is the Gauss-Bonnet term)  \cite{fG} and $f(G,T)$ \cite{fGT} etc., are among the most attractive models of modified theories (for further reviews on
such models, see, for
instance,~\cite{R-NO-CF-CD}). Harko \textit{et al.} \cite{harko1} introduced a notion of $f(R,T)$ theory by introducing $T$ in the well-known theory of $f(R)$ gravity. This quantity $T$ may be considered by exotic
imperfect fluids or quantum effects. They presented dynamical equations and the associated equations of motion for a test particle. Houndjo \cite{hound1} did reconstruction in order to solve cosmological issues in
this gravity and found some models that could be useful to understand matter dominated eras
of our universe. Jamil \emph{et al.} \cite{ya11} did the same process and found some well-consistent outcomes with low red-shifts surveys.

Adhav \cite{3i} studied the homogeneous and anisotropic cosmic model with the help of a constant deceleration parameter and presented a few analytical $f(R,T)$ solutions under some conditions. Shabani and Farhoudi \cite{9i} examined the cosmological solutions of $f(R,T)$ gravity for a perfect fluid using spatially Friedmann-Lema\^{i}tre-Robertson-Walker universe. In order to
simplify equations, they presented some dimensionless parameters and
variables. Baffou \emph{et al.} \cite{4i} analyzed the stability of power-law models with the help of de-Sitter
cosmic models against linear perturbations. They concluded that these models could be taken as an
efficient candidate for DE. Das and Ali \cite{6i} elaborated the anisotropic and homogeneous axially
symmetric Bianchi type-I bulk viscous cosmological models using time-varying
cosmological and gravitational constant. By using the
Hubble parameter, they solved the field equations and discussed the kinematical and physical properties of the
models. Kiran and Reddy \cite{7i} investigated the Bianchi type-III in
the presence of viscous fluid and concluded that this model does not
exist in $f(R,T)$ gravity. Momeni \emph{et al.} \cite{5i} analyzed the Noether symmetry problem for two
types of modified theories. First is mimetic $f(R)$ and second is
a non-minimally coupled model which is known as  $f(R,T)$.
Pankaj and Singh \cite{8i} discussed the viscous
cosmology with matter creation under $f(R,T)$ gravity. Sun and Huang \cite{11i} studied the issues of an isotropic and homogeneous universe
under modified $f(R,T)$ gravity with non-minimal coupling. Moraes \emph{et al.} \cite{13i} studied the hydrostatic equilibrium conditions of compact objects by relating their pressure and density through an equation of state.

A gravastar (Gravitational Vacuum Star) is an astronomical substance
hypothesized as a substitute to the black hole. The conception of
gravastar was established from the theory of Mazur and Mottola,
\cite{1,2}. By increasing the idea of Bose-Einstein,
this new form of solution was introduced as a consequence of
gravitational collapse. Such kind of model is hypothesized to contain no event horizon. Though gravastar may look identical to a black hole, there are
certain experiments, such as X-rays radiations by
infalling matter, that may allow
to discriminate them. As such, gravastars are of interest from a
purely theoretical perspective. Such kind of stellar structures could be used to explain the role of DE
in the accelerating expansion of the universe. These could be helpful
to explain why some galaxies have a high or low DM concentrations. The gravastars could be described with the help of three different zones, in which
I is the interior region ($0\leq r,~r<r_1$), II is the intermediate thin shell, with $r_1<r,~r<r_2$, while III is an
exterior region ($r_2<r$). It is so happened that
in the region I, the isotropic
pressure produces a force of repulsion over the intermediate thin
shell, which is equal to -$\rho$ (where $\rho$ is the energy density).
This intermediate thin shell is supposed to be supported by a
fluid pressure and ultrarelativistic plasma. However, the region III can be represented by
the vacuum solution of the field equations. The pressure has zero value at this zone. It contains thermodynamically stable solution and maximum entropy under small fluctuations \cite{1,2}.

Visser \cite{1gh} developed a simple mathematical model for the description of Mazur-Mottola scenario and described the
stability of gravastars after exploring some realistic values of EoS parameter. Cattoen \emph{et al.} \cite{2gh} extended their results for the case of anisotropic gravastar structures. They calculated the anisotropic factor $\Delta$ from the equations of motion for the spherically symmetric spacetime and analyzed the inclusion of pressure anisotropy could be useful to support relatively high compact gravastars. Based upon the ranges of the parameters involved, Carter \cite{3gh} studied the stability of the gravastar and checked the existence of thin shell. He used Israel junction condition in order to join de-Sitter spacetime (interior) with a Reissner-Nordstr\"{o}m (RN) exterior metric. He also analyzed the role of EoS parameter in the modeling of gravastar structures.

Horvat \emph{et al.} \cite{4gh}  presented two different theoretical models of the gravastars in the presence of an electromagnetic field. After joining the interior metric with an appropriate exterior vacuum geometry, they obtained viability constraints for the stability of gravastar though dominant energy conditions. They also studied the effects of electromagnetic field on the formulations as well as graphical representations of EoS, the speed of sound and the surface red-shift. Rahaman \emph{et al.} \cite{5gh} discussed the existence of charged gravastar in an environment of
(2+1)-dimensional spacetime. They studied various physical
properties, like, length and energy within the thin shell and
entropy of the charged gravastars and claimed that their solutions are nonsingular that could present
a viable alternative to the black hole.

De Felice \emph{et al.} \cite{13} found exact solutions
of the spherically symmetric spacetimes in the presence of electric
charge and also compared their results with the existing black holes
models. It can be concluded that one can mollify the phenomenon of gravitational collapse to a great extent
in the presence of a electric charge. Yousaf and Bhatti \cite{m1} investigated the modeling of relativistic structures in the presence of electromagnetic field. They concluded that the electric charge has greatly weakens the influence of modified gravity leading to produce a repulsive field. Turimov \emph{et al.} \cite{turi1} studied the slowly rotating magnetized gravastars in the presence of electromagnetic field.

Rahaman \emph{et al.} \cite{16} studied the 3-dimensional neutral
spherically symmetry model of gravitational vacuum star whose
exterior region is elaborated by \emph{Ba\~{n}ados-Teitelboim-Zanelli metric}. He presented a
non-singular and stable model and discussed various physical
features, for example, length, energy conditions, entropy, and
junction conditions of the spherical distribution by using static
spherically symmetric matter distribution as an interior spacetime.
Lobo and Garattini \cite{17} performed linearized
stability analysis with noncommutative geometry of gravastars and
investigated few exact solutions of gravastar after exploring their
physical features and characteristics.

Usmani \emph{et al.} \cite{19} studied charged gravastar
experiencing conformal motion and elaborated the dynamics of the
formation of thin shell and the entropy of the system. Herrera and de Le\'{o}n \cite{21} discussed the role of
anisotropic/isotropic pressure on charged spheres and found some
exact solutions of the non-linear field equations by assuming
spherical symmetry spacetime. The same authors \cite{22} also
analyzed the dynamics of anisotropic spheres by introducing one
parameter group of conformal motions. They inferred that on the
boundary of matter, all of their calculated solutions can exist on
the Schwarzschild exterior metric. Esculpi and Aloma \cite{23}
studied the conformal motion of the charged fluid sphere with linear
equation of state (EoS). They also discussed dynamical stability
analysis of the relativistic structures. The process of dynamical
instability \cite{dyn1} and the regularity of certain physical
quantities \cite{inho1} on the surface of evolving matter
distribution are also discussed in literature. Ray and Das \cite{24}
discussed the electromagnetic mass model with the help of conformal
killing vector. By considering the existence of a one parameter
group, the corresponding conformal motion have been described for
the charged strange quark star model.

This paper is devoted to understand the existence of gravastar
under spherically symmetric spacetime in the realm of
Maxwell-$f(R,T)$ gravity. The paper is arranged as follows. In
section \textbf{2}, we shall describe the basic frame work of
$f(R,T)$ theory and their conservation equation. Section \textbf{3}
is based on the formulation of field/conservation equations and
gravitational mass of the static spherically symmetric manifold. In section \textbf{4}, we shall
calculate the mass of thin shell by using certain
matching conditions. The section \textbf{5} is aimed to discuss the
effects of charge on the various physical features of gravastar.
Finally, we summarize our findings.

\section{$f(R,T)$ Gravity}

The action of the $f(R,T)$ theory is given as follows
\begin{eqnarray}\label{1}
\textbf{S}=\frac{1}{16\pi}\int{f(R,\textit{T})}\sqrt{-g}~d^4x+\int\textit{\L}_m\sqrt{-g}~d^4x,
\end{eqnarray}
where $f(R,T)$ is the arbitrary function of $R$ and $T$ with $R$ as
the Ricci scalar and $T$ as the trace of the energy-momentum tensor,
$\textit{\L}_m$ is the Lagrangian matter density, and $g$ is the
determinant of metric tensor $g_{\zeta\eta}$. In this paper, we have
consider $G = c = 1$. By varying the action of $f(R,T)$ with respect
to the metric $g_{\zeta\eta}$, we can deduce the field equations of
$f(R,T)$ of gravity as
\begin{eqnarray}\label{2}
{f_{R}(R,\textit{T})}R_{\zeta\eta}-\frac{1}{2}{f(R,\textit{T})}g_{\zeta\eta}
-(\nabla_\zeta\nabla_\eta-g_{\zeta\eta}\Box){f_{R}(R,\textit{T})}
+{f_{T}(R,\textit{T})}(T_{\zeta\eta}+\Theta_{\zeta\eta})=8\pi\textit{T}g_{\zeta\eta},
\end{eqnarray}
where $f_{R}(R,T)$ is the derivative of generic function $f$ with
respect to the Ricci scalar $R$, $f_{T}(R,T)$ is the derivative of
generic function with respect to trace of the energy-momentum tensor
$T$, $\Box$ is the product of contravariant and covariant
derivative,
\begin{align}\label{3a}
\Theta_{\zeta\eta}=g_{\zeta\eta}\frac{\partial
T_{\zeta\eta}}{\partial g_{\zeta\eta}}
\end{align}
where the stress-energy tensor is defined below
\begin{eqnarray}\label{3}
T_{\zeta\eta}= g_{\zeta\eta}\textit{\L}_m -2\frac{\partial
\textit{\L}_m}{\partial g^{\zeta\eta}}.
\end{eqnarray}
We assume perfect fluid as the energy-momentum tensor
\begin{eqnarray}\label{5}
T_{\zeta\eta}=(\rho+p) U_{\zeta}U_{\eta}- pg_{\zeta\eta},
\end{eqnarray}
where $\rho$ is the density, $p$ is the pressure and $U_{\zeta}$ is
the 4-velocity vector. It is interesting to mention that here the
equations of motion is dependant on the role of fluid distribution.
Therefore, one can take specific equations of motion by choosing
$\textit{\L}_m$. In literature, researchers chose $\textit{\L}_m=p$
and $\textit{\L}_m=\rho$ \cite{harko1, 4i, hound1, bam1, ya13}. Here, we
are interested to study the charged isotropic fluid, therefore we
shall take $\textit{\L}_m$=$-\left(p+\mathcal{F}\right)$, where
$\mathcal{F}$ shows the contribution of electromagnetic field and is
defined through Maxwell tensor $(F_{\alpha\beta})$ as
$\mathcal{F}=\frac{1}{16\pi}F_{\alpha\beta}F_{\gamma\sigma}g^{\alpha\gamma}g^{\beta\sigma}$.
The Maxwell field tensor is defined through four potential
$\phi_\alpha$ as
$F_{\alpha\beta}=\phi_{\beta,\alpha}-\phi_{\alpha,\beta}$. The
Maxwell equations of motion are
\begin{equation}\label{5}
F^{\xi\beta}_{~~;\beta}={K}_{0}J^{\xi},\quad
F_{[\xi\beta;\gamma]}=0,
\end{equation}
in which $J^\xi$ is the four current and $K_0$ indicates the
permeability of the magnetic field. In view of this scenario,
Eq.\eqref{3a} reduces to
$$\Theta_{\zeta\eta}=-2T_{\zeta\eta}-p
g_{\zeta\eta}-\mathcal{F}g_{\zeta\eta}.$$ In this work, we use
functional form of $f(R,T)$ = $2\chi T+R$. Substituting this
relation in Eq.\eqref{2}, we get
\begin{eqnarray}\label{6}
G_{\zeta\eta}=8\pi(T_{\zeta\eta}+E_{\zeta\eta})+\chi
Tg_{\zeta\eta}+2\chi(T_{\zeta\eta}+E_{\zeta\eta}+pg_{\zeta\eta}+\mathcal{F}g_{\zeta\eta}),
\end{eqnarray}
where $G_{\zeta\eta}$ is the Einstein tensor and and $E_{\zeta\eta}$
is the energy momentum tensor for an electromagnetic field and is
given by \cite{m34}
\begin{equation}\label{4n}
E_{\xi\beta}=\frac{1}{4\pi}\left(F^{\gamma}_{\xi}F_{\beta\gamma}
-\frac{1}{4}F^{\gamma\delta}F_{\gamma\delta}g_{\alpha\xi}\right).
\end{equation}
In this paper, we consider a scenario in which the system is
evolving by keeping the charged particles at the state of rest. This
will give zero contribution to the magnetic field. Therefore,
\begin{equation*}\label{5a}
\phi_{\xi}={\Phi}(t,r){\delta^{0}_{\xi}},\quad
J^{\xi}={K_1}(t,r)V^{\xi},
\end{equation*}
where $\Phi$ represents the corresponding scalar potential and the
and $K_1$ indicates charge density.

To understand $f(R,T)$ theory as an appropriate gravitational theory, one must
consider a viable and effective distribution of $f(R,T)$ function. Besides its physical consistency
with the observations of current cosmic acceleration, it should pass the stability tests and should meet
the viability requirements from solar and terrestrial
static/non-static systems. Usually, the $f(R,T)$ models are presented in the following different ways:
\begin{enumerate}
  \item $f(R,T)=R+2g(T)$. Such kind of selection in the geometric part of the Lagrangian describes cosmological constant $\Lambda$ as a time-dependent entity and hence represents the $\Lambda$CDM model..
  \item $f(R,T)=f_1(R)+f_2(T)$. This type of choice corresponds to the minimal knowledge to understand modified relativistic dynamics. This could be regarded as the corrections to the notable $f(R)$ theory. By considering any linear combination of $f_2$, various distinct results can be obtained from the choices of $f(R)$ function.
  \item $f(R,T)=f_1(R)+f_2(T)f_3(R)$. This explicitly describes non-minimal coupled matter-geometry theory of gravity. The comparison of results found from this selection may be different from the minimal interacting choices.
\end{enumerate}
The criteria for understanding the viability of $f_1(R)$ model are as follows
\begin{itemize}
\item For positive value of $f_{1R}(R)$ with $R>\bf{R}$; where $\bf{R}$
represents today's choice of the Ricci scalar. This condition is necessary to prevent the
appearances of a ghost state. Ghosts often appear which notify that
DE is responsible for cosmic acceleration under modified
gravity theories. This state could be induced by a mysterious force which
creates a repulsion between the supermassive or massive stellar object.
For retaining the attractive feature of gravity, the constraint
should maintain a positive sign with a consistent gravitational
constant, $G_{eff}=G/f_{1R}$.
\item The positive value of $f_{1RR}(R)$ with $R>\bf{R}$.
This requirement is introduced for making the evolving system not to conceive situations in which tachyons appears. A hypothetical object that could move faster than the speed of light is known as
tachyon.
\end{itemize}
If a model of $f_1(R)$ does not fulfill those conditions, it would
not be considered viable. Haghani \emph{et al.} \cite{zg1} as well as Odintsov and D. S\'{a}ez-G\'{o}mez \cite{zg2} proposed that Dolgov-Kawasaki instability in $f(R,T)$ gravity requires a similar
sort of limitations as in $f(R)$ gravity and one needs to satisfy $1+f_T>0$ with
$G_{eff}>0$. Thus, in the realm of $f(R,T)$ models, the following conditions should be fulfilled
\begin{eqnarray*}
f_R>0,~~~~~1+f_T>0,~~~~~f_{RR}>0,~~~~~R>\bf{R}
\end{eqnarray*}
Thus, throughout in our paper,  we assume that $1+2\chi>0$. One thing which
must be taken into account is that the divergence of the energy-momentum tensor
is not zero in $f(R,T)$ gravity and is defined as
\begin{eqnarray}\label{non1}
\nabla^\zeta T_{\zeta\eta}=\frac{f_{T}}{8\pi-f_{T}}[(T_{\zeta\eta}
+\Theta_{\zeta\eta})\nabla^\zeta\ln{f_T}+\nabla^\zeta\Theta_{\zeta\eta}
-\frac{1}{2}g_{\zeta\eta}\nabla^\zeta T].
\end{eqnarray}
The non-zero value of divergence of energy momentum tensor causes the breaking of all equivalence principle in $f(R,T)$
gravity. According to the weak equivalence principle,``\textit{All test
particles in a given gravitational field will undergo the same
acceleration, independent of their properties, including their rest
mass."}. In this modified theory, the equation of motion is based on
those feature of the particle that ate thermodynamic in nature, e.g.,
pressure, energy density, etc. Further, the strong equivalence principle
states that ``\textit{The gravitational motion of a small test body
depends only on its initial position and velocity, and not on its
configuration."} \cite{class1} .This principle also does not hold in $f(R,T)$ theory thus causing the particles to experience
non-geodesic motion along the world lines. In the background of quantum theory, one can relate the non-zero divergence of the effective energy-momentum tensor with the violation of energy conservation in the scattering phenomenon. In this theory, the energy non-conservation can cause an energy flow between the four-dimensional spacetime and a compact extra-dimensional metric \cite{nn1}. It is worth notable that the constraint $f(T)=0$ in Eq.\eqref{non1} would reduce our dynamics to that of $f(R)$ gravity. One can write Eq.\eqref{non1} as follows
\begin{eqnarray}\label{7}
\nabla^\zeta
T_{\zeta\eta}=\frac{-2\chi}{8\pi+2\chi}\left[\nabla^\zeta(pg_{\zeta\eta})
+\nabla^\zeta(\mathcal{F}g_{\zeta\eta})+\frac{1}{2}g_{\zeta\eta}\nabla^\zeta
T\right].
\end{eqnarray}
It is worthy to stress that Eq.\eqref{non1} corresponds to a general $f(T)$ part,
while Eq.\eqref{7} describes the covariant divergence of stress-energy tensor with a linear contribution
on $f(T)$ model.

\section{Spherically Symmetric Spacetime Models}

This section is devoted to explore modified equations of motion,
including field and conservation laws. By simultaneous solving these
equation, we evaluate evolution equation, After using a specific
combination of EoS, we shall evaluate the value of the corresponding
scale factors that would eventually leads to gravitational mass of
the relativistic structure. We consider irrotational static form of
the spherically spacetime as follows
\begin{eqnarray}\label{8}
ds^2=e^{\nu(r)}dt^2-e^{\lambda(r)}dr^2-r^2(d\theta^2+\sin\theta^2d\phi^2).
\end{eqnarray}
The nonzero components of the Einstein tensor for above equation are
\begin{align}\label{9}
G^{00}&=\frac{\lambda'
r+e^{\lambda}-1}{r^2e^{\nu}e^{\lambda}},\\\label{10}
G^{11}&=\frac{\nu' r-e^{\lambda}+1}{r^2e^{\lambda^2}},\\\label{11}
G^{22}&=\frac{-2\lambda'-2\nu'+(2\nu''+\nu'^2-\nu'\lambda')r}{4r^3e^{\lambda}}.
\end{align}
The nonzero components of Eq.\eqref{4n} are given as
\begin{eqnarray}\label{12}
E_{00}=2\pi E^2e^{\nu},~~~ E_{11}=2\pi E^2e^{\lambda},~~~E_{22}=2\pi
E^2r^2,
\end{eqnarray}
where $E$ is the electric intensity, which is defined via electric
charge ($q$) as
\begin{eqnarray*}
E=\frac{q}{4\pi r^2}.
\end{eqnarray*}
After using Eqs.\eqref{9}-\eqref{12} in Eq.\eqref{5}, we get
\begin{align}\label{13}
&{\lambda'r+e^{\lambda}-1}=r^2e^{\lambda}\left[8\pi\rho+\chi(3\rho-p)+\frac{q^2}{r^4}
+\frac{\chi}{r^4}\left(\frac{1}{4\pi}+1\right)
q^2\right], \\\label{14}
&{-\nu'r+e^{\lambda}-1}=r^2e^{\lambda}\left[-8\pi
p+\chi(\rho-3p)+\frac{q^2}{r^4}+\frac{\chi}{r^4}\left(\frac{1}{4\pi}+1\right)
q^2\right], \\\label{15}
&-\frac{r}{2}(\nu'-\lambda')+\frac{r^2}{4}(\nu'\lambda'-2\nu''-\nu'^2)=r^2e^{\lambda}\left[-8\pi
p+\chi(\rho-3p)-\frac{q^2}{r^4}+\frac{\chi}{r^4}\left(-\frac{1}{4\pi}+1\right)
q^2\right].
\end{align}
The hydrostatic equilibrium condition can be evaluated with the help
of conservation law as
\begin{align}\label{16}
\frac{\nu'}{2}(\rho+p)+\frac{dp}{dr}+\frac{3q^2}{8\pi r^5}
+\frac{\chi}{4\pi+\chi}\left[\frac{2q^2}{r^5}+\frac{1}{2}(\rho'-p')\right]=0.
\end{align}
With the help of Misner-Sharp formula \cite{ya32} and $G_{00}$ component of the Einstein tensor, the
corresponding component of line element $g_{00}$ becomes
\begin{eqnarray}\label{17}
e^{-\lambda}=1-\frac{2m}{r}-\chi\left(\rho-\frac{p}{3}\right)r^2+\frac{2q^2}{r^4}
+\frac{\chi}{r^4}\left(\frac{1}{4\pi}+1\right)
q^2.
\end{eqnarray}
Using Eq.\eqref{14} in  Eq.\eqref{16}, we get
\begin{eqnarray}\label{18}
\frac{dp}{dr}=\frac{-\frac{2q^2\chi}{r^5(4\pi+\chi)}-\frac{3q^2}{8\pi
r^5}-\frac{\nu'}{2}(\rho+p)}{\left[1+\frac{\chi}{2(4\pi+\chi)}\left(1-\frac{d\rho}{dp}\right)\right]},
\end{eqnarray}
where
\begin{eqnarray*}
\nu'=\frac{r\left[8\pi
p-\chi(\rho-3p)-\frac{q^2}{r^4}-\frac{\chi}{r^4}\left(\frac{1}{4\pi}+1\right)q^2\right]
+\frac{1}{r}\left[\frac{2m}{r}+\chi\left(\rho-\frac{p}{3}\right)-
\frac{2q^2}{r^4}-\frac{2\chi}{r^4}\left(\frac{1}{4\pi}+1\right)
q^2\right]}{\left[1+\frac{2q^2}{r^4}-\frac{2m}{r}-\chi\left(\rho
-\frac{p}{3}\right)r^2+\frac{2\chi}{r^4}\left(\frac{1}{4\pi}+1\right)
q^2\right]}.
\end{eqnarray*}
Gravastars \cite{1,2} consists of three regions characterized by an
EoS $p=\omega\rho$, where $\omega$ is constant. Here, we assume that
the interior region is filled with an enigmatic gravitational
source. The corresponding EoS for dark energy model is given as
\begin{eqnarray}\label{19}
p=-\rho,~\quad \textrm{with}~\omega=-1.
\end{eqnarray}
Using, $\rho=\rho_0$ (constant) in Eq.\eqref{19}, we get
\begin{eqnarray}\label{20}
p=-\rho_0.
\end{eqnarray}
After using Eq.\eqref{19} in  Eq.\eqref{13}, it follows that
\begin{eqnarray}\label{21}
e^{-\lambda}=1-\frac{4r^2\rho_0}{3}(2\pi+\chi)+\frac{q^2}{r^2}
+\frac{\chi}{r^2}\left(\frac{1}{4\pi}+1\right)
q^2+\frac{H}{r},
\end{eqnarray}
where $H$ is an integration constant, whose value, after applying
the regularity condition is found to be zero. Therefore,
Eq.\eqref{22} becomes
\begin{eqnarray}\label{22}
e^{-\lambda}=1-\frac{4r^2\rho_0}{3}(2\pi+\chi)+\frac{q^2}{r^2}
+\frac{\chi}{r^2}\left(\frac{1}{4\pi}+1\right)
q^2.
\end{eqnarray}
Substituting an EoS in Eqs.\eqref{13} and \eqref{14}, we have
\begin{eqnarray}\label{23}
e^{-\lambda}=Ie^{\nu},
\end{eqnarray}
where $I$ is an integration constant. The gravitational mass $M(D)$
can be found as follows
\begin{eqnarray}\label{24}
M(D)=\int^{r=D}_04\pi\left(\rho_0+\frac{q^2}{2r^2}\right)r^2dr=2\pi
D\left(\frac{2}{3}D^2+q^2\right),
\end{eqnarray}
where $\rho_0$ is the constant density. Equation \eqref{24} describes
that the interior gravitational mass and
radius of the stellar system are directly proportional to each other. This is the characteristic feature of the stellar
compact object. Furthermore, the above equation also states the substantial dependence of
$M$ on the specific value $r=D$ in the presence of electric charge. This integral becomes improper on substituting $r=\infty$. However, this choice is not realistic as one can not consider the infinite radius of the stellar body.

\section{Intermediate Shell of the Charged Gravastar}

\begin{center}\begin{figure}\centering
\epsfig{file= 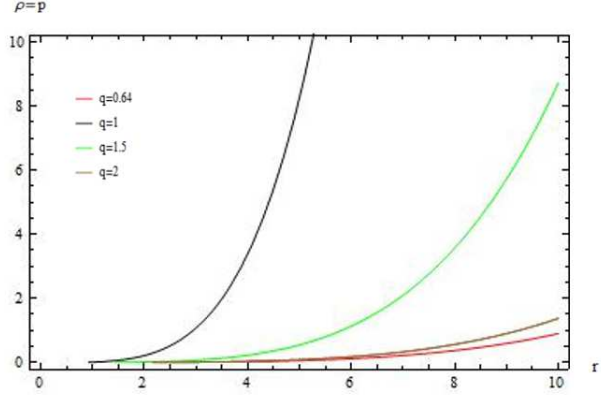,width=0.5\linewidth}\caption{\small{Plot of
pressure $p$ within the shell with respect to radius $r$(km) with different charges.}}\label{a1}
\end{figure}
\end{center}

In this section, we tend to discuss the effect of electromagnetic
charge on the formulation of intermediate shell of the corresponding
gravastars. We shall also explore the smooth matching conditions for
the joining of interior and exterior manifolds of the gravastar
structures by using Darmois-Israel formalism. For this purpose, we
assume that the intermediate shell is formed by an ultrarelativistic
fluid with non-vacuum background with an equation of state $p=-\rho$.
It is hard to calculate the solution of the cumbersome set of field
equations in the non-vacuum region. To avoid this query, we will use
some approximation and find analytical solution, i.e.,
$0<e^{-\lambda}\ll1$. By solving field Eqs.\eqref{13}-\eqref{15}
under EoS, we end up with the following two equations as
\begin{align}\label{25}
&\frac{de^{-\lambda}}{dr}=-\frac{2q^2}{r^3}-\frac{2\chi}{r^3}\left(\frac{1}{4\pi}
+1\right)q^2+\frac{2}{r},\\\label{26}
&\left(\frac{3}{2r}+\frac{\nu'}{4}\right)\frac{de^{-\lambda}}{dr}=-\frac{2\chi
q^2}{r^4}+\frac{1}{r^2}.
\end{align}
Integrating Eq.\eqref{25}, we get
\begin{eqnarray}\label{27}
e^{-\lambda}=\frac{q^2}{r^2}+\frac{2\chi}{r^2}\left(\frac{1}{4\pi}+1\right)q^2+2lnr+B,
\end{eqnarray}
where $B$ is an integration constant and $r$ is the radius belonging
to $D\ll r \ll D+\epsilon$, under $\epsilon\ll1$. To get analytical
values of pressure and radius of thin shell, we use Eqs.\eqref{25}
and \eqref{26} in Eq.\eqref{16}, we get the behavior of pressure
with respect to radius, which are shown in Fig.\eqref{a1}.

In order to discuss the structure of gravastar, we take static
Schwarzschild spacetime as an exterior geometry given as follows,
\begin{eqnarray}\label{28}
ds^2=\left(1-\frac{2M}{r}+\frac{q^2}{D^2}\right)dt^2-\left(1-\frac{2M}{r}
+\frac{q^2}{D^2}\right)^{-1}dr^2-r^2(d\theta^2+\sin\theta^2d\phi^2).
\end{eqnarray}
Darmois  \cite{25} and Israel \cite{26} introduced conditions for the matching of interior and exterior geometries over the surface. The metric coefficients are continuous at the junction
surface $(\Sigma)$, i.e., their derivatives might not be continuous
at interior surfaces. The surface tension and surface stress energy
of the joining surface $S$ may be resolved from the discontinuity of
the extrinsic curvature of $S$ at $r=D$. The field equation of
intrinsic surface is defined by Lanczos equation as
\begin{eqnarray}\label{29}
S_\alpha^\beta=-\frac{1}{8\pi}(\Upsilon_\alpha^\beta-\delta_\beta^\alpha
\Upsilon_\kappa^\kappa),
\end{eqnarray}
where $S_{ji}$ is the stress-energy tensor for surface,
$\Upsilon_{\alpha\beta}=\eta_{\alpha\beta}^{+}-\eta_{\alpha\beta}^-$
tells the extrinsic curvatures or second fundamental forms and $(+)$
sign indicates the interior surface while $(-)$ sign indicates the
exterior surface. The second fundamental forms connects interior and
exterior surfaces of the thin shell and are defined as follows
\begin{eqnarray}\label{30}
\eta_{\mu\nu}^{\pm}=-n_i^{\pm}\left[\frac{\partial^2x_i}{\partial\xi^\mu
\partial\xi^\nu}+\Gamma_{\gamma\delta}^i\frac{\partial
x^\gamma}{\partial\xi^\mu}\frac{\partial
x^\delta}{\partial\xi^\nu}\right]_\Sigma,
\end{eqnarray}
where $\xi^\mu$ represents the coordinate of intrinsic metric and
$n_i^{\pm}$ describes the unit normals on the surface of gravastar.
\begin{eqnarray}\label{31}
n_i^{\pm}=\pm\left|g^{\alpha\beta}\frac{\partial f(r)}{\partial
x^\alpha}\frac{\partial f(r)}{\partial
x^\beta}\right|^{-\frac{1}{2}}\frac{\partial f(r)}{\partial x^i},
\quad{n_jn^j=1},
\end{eqnarray}
where $f(r)$ illustrates the coordinate of exterior exterior metric.
Using the Lanczos equations, we can get surface energy density
($\varphi$) and surface pressure $\psi$ as
\begin{align}\label{32}
\varphi&=-\frac{1}{4\pi D}[\sqrt{f(r)}]_-^+,\\\label{33}
\psi&=-\frac{\varphi}{2}+\frac{1}{16\pi}\left[\frac{f(r)'}{\sqrt{f(r)}}\right]_-^+.
\end{align}
Making use of Eqs.\eqref{32} and \eqref{33}, we get
\begin{align}\label{34}
\varphi&=-\frac{1}{4\pi
D}\left[\sqrt{1-\frac{2M}{D}+\frac{q^2}{D^2}}-\sqrt{1-\frac{4D^2\rho_0}{3}(2\pi
+\chi)+\frac{q^2}{D^2}+\frac{\chi}{D^2}\left(\frac{1}{4\pi}+1\right)
q^2}\right],\\\label{35} \psi&=\frac{1}{8\pi
D}\left[\frac{\left(1-\frac{M}{D}\right)}{\sqrt{1-\frac{2M}{D}+\frac{q^2}{D^2}}}
-\frac{\left[1-\frac{8D^2\rho_0}{3}(2\pi+\chi)+
\frac{q^2}{2D^2}+\frac{\chi}{2D^2}\left(\frac{1}{4\pi}+1\right)
q^2\right]}{\sqrt{1-\frac{4D^2\rho_0}{3}(2\pi+\chi)+\frac{q^2}{D^2}+\frac{\chi}
{D^2}\left(\frac{1}{4\pi}+1\right)
q^2}}\right].
\end{align}
One can find mass of the intermediate thin shell by using areal
density as
\begin{eqnarray}\label{36}
m_s=4\pi
D^2\varphi=-D\left[\sqrt{1-\frac{2M}{D}+\frac{q^2}{D^2}}-\sqrt{1-\frac{4D^2\rho_0}
{3}(2\pi+\chi)+\frac{q^2}{D^2}+\frac{\chi}{D^2}\left(\frac{1}{4\pi}+1\right)
q^2}\right],
\end{eqnarray}
where
\begin{align*}
M&=\frac{m_s}{D}{\sqrt{1-\frac{4D^2\rho_0}{3}(2\pi+\chi)+\frac{q^2}{D^2}
+\frac{\chi}{D^2}\left(\frac{1}{4\pi}+1\right)
q^2}}-\frac{m_s^2}{2D}+\frac{2\rho_oD^3}{3}(2\pi+\chi)\\\nonumber
&-\frac{\chi}{2D}\left(\frac{1}{4\pi}+1\right) q^2,
\end{align*}
represents the total mass of the gravitational vacuum star with $m_s=m$. It can be noticed that one can calculate the value of $M$, once the mass of intermediate thin shell ($m_s$),
 the radial distance ($D$) as well as the value of $f(R,T)$ correction term ($\chi$) or electric charge ($q$) are known.
It also indicates that the physical quantities $m_s,~D$ and $q$ have dominated their influence over the $f(R,T)$ dark source terms. This is because one can diminish the role of $\chi$ by substituting zero to $m_s,~D$ and $q$. This situation could be different, if one consider the Palatini $f(R,T)$ gravity \cite{fRtplat1} instead of metric $f(R,T)$ gravity \cite{harko1}.

\begin{center}\begin{figure}\centering
\epsfig{file= 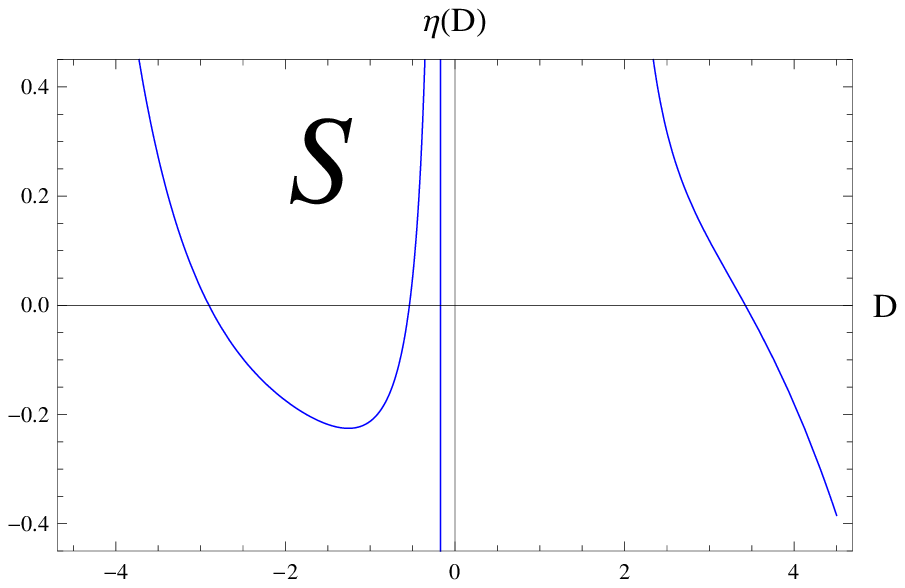,width=0.45\linewidth}\epsfig{file=
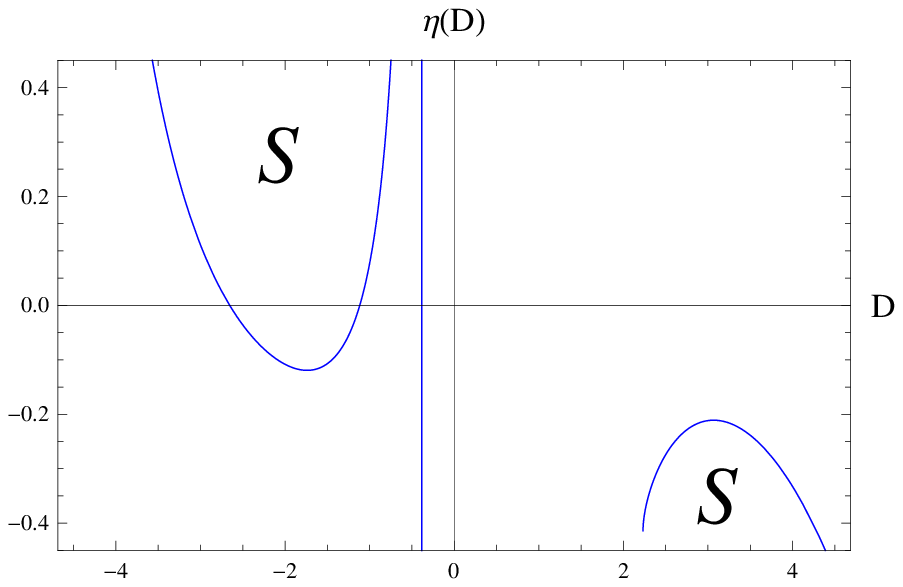,width=0.45\linewidth}\caption{\small{Stability regions of
the charged gravastar in terms of
$\eta=\frac{\acute{\psi}}{\acute{\varphi}}$. We have chosen $\chi=0.2$,
$M=1.2$, $\rho=0.002$} at $q=1$ and $q=1.5$.}\label{b1a}
\end{figure}
\end{center}

\begin{center}\begin{figure}\centering
\epsfig{file= 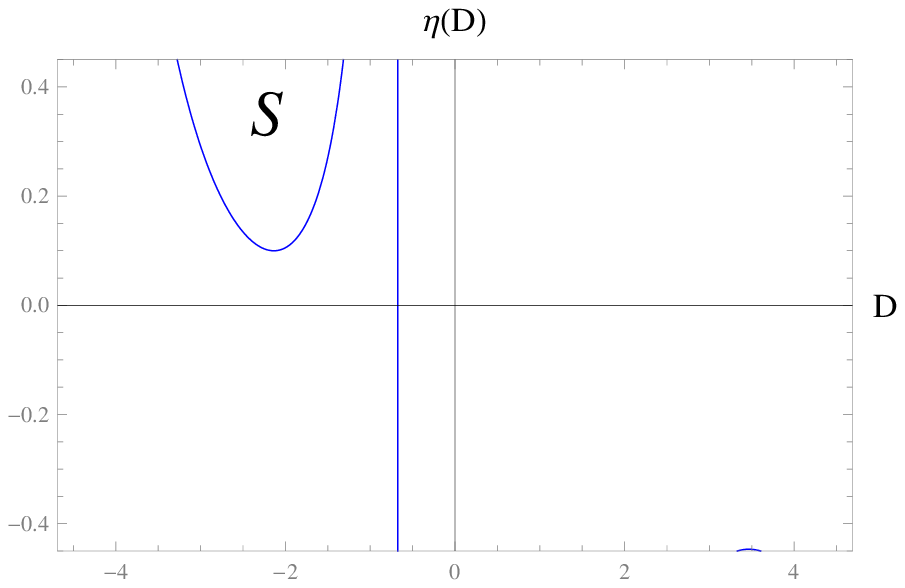,width=0.45\linewidth}\epsfig{file=
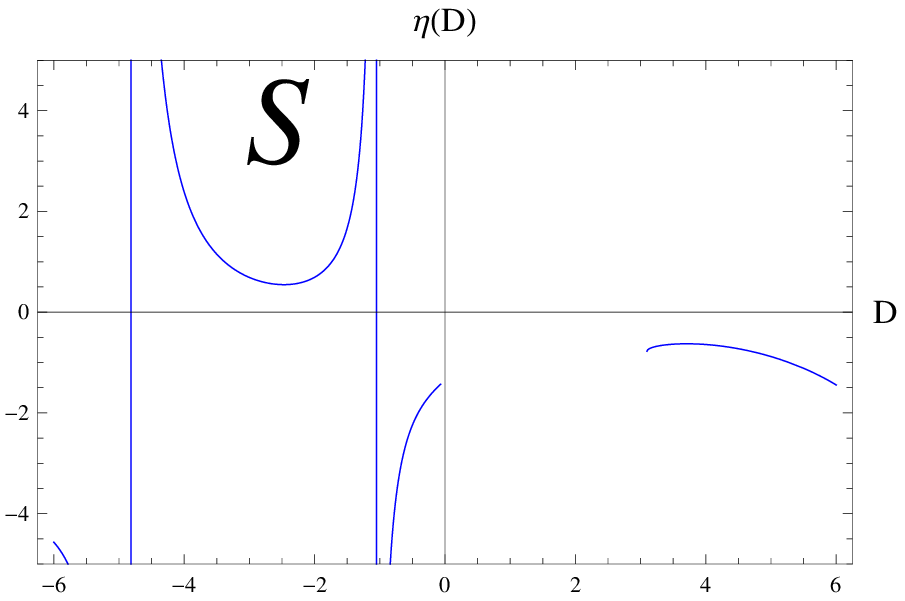,width=0.45\linewidth}\caption{\small{Stability regions of
the charged gravastar in terms of
$\eta=\frac{\acute{\psi}}{\acute{\varphi}}$. We have chosen $\chi=0.2$,
$M=1.2$, $\rho=0.002$} at $q=2$ and $q=2.5$.}\label{b1b}
\end{figure}
\end{center}

It will be very useful to understand the stability of gravastars by defining a parameter ($\eta$) as the ratio of the derivatives of $\psi$ and $\varphi$ as follows
$$\left.\eta(a)=\frac{\psi'(a)}{\varphi'(a)}\right|_{r=a_0}.$$
The stability regions can be explored by analyzing the behavior of $\eta$ as a function of $r=a_0$. \"{O}vg\"{u}n
\emph{et al.} \cite{ovgun} considered static form of the spherically symmetric spacetime and analyzed the the stability of a charged thin-shell gravastar with the help of a similar parameter as defined above. We have investigated the stable regimes of gravastars with specific choices of parameters involved. The letter $S$ in Figs.\ref{b1a} and \ref{b1b} describe the stable epochs of spherically symmetric gravastar structures.

\section{Some Features of Gravastars}

This section is devoted to examine the impact of electromagnetic
field on different physical features of gravastar. In this context,
we shall calculate proper length of the thin shell as well as energy
of relativistic structure. After examining the entropy of
gravastars, the role of EoS parameter will be analyzed on the
dynamical formulation of gravastars. We shall also describe our
results by drawing various diagrams and graphs.

\subsection{Proper Length of the Thin Shell}

\begin{center}\begin{figure}\centering
\epsfig{file= 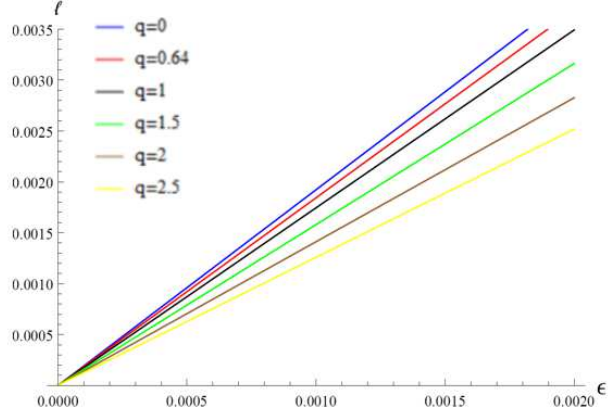,width=0.51\linewidth}\caption{\small{Plot of
length $\ell$(km) of the shell with thickness of the shell
$\epsilon$(km). Fixing  $\chi=1$, $C=0.00006$, $D=1$.}}\label{b1}
\end{figure}\end{center}

In this subsection we shall consider $r=D$ for describing the radius
of an interior region, while $r=D+\epsilon$ (with $\epsilon \ll 1$)
and $\epsilon$ represent the radius of exterior region and the
thickness of the intermediate thin shell, respectively. The proper
thickness between two surfaces can be described mathematically as
\begin{eqnarray}\label{37}
\ell=\int^{D+\epsilon}_D\sqrt{e^\lambda}dr=\int^{D+\epsilon}_D
\frac{1}{\sqrt{\frac{q^2}{r^2}+\frac{2\chi}{r^2}(\frac{1}{4\pi}+1)
q^2+2lnr+C}}dr,
\end{eqnarray}
The analytic solution of the above expression with Maxwell-$f(R,T)$
gravity corrections is not possible. We shall solve it by numerical
method and examine the behavior of charge. The behavior of length of
the shell vs its thickness has been shown in Fig.\eqref{b1}.

\subsection{Energy of the Charged Gravastar}

\begin{center}\begin{figure}\centering
\epsfig{file= 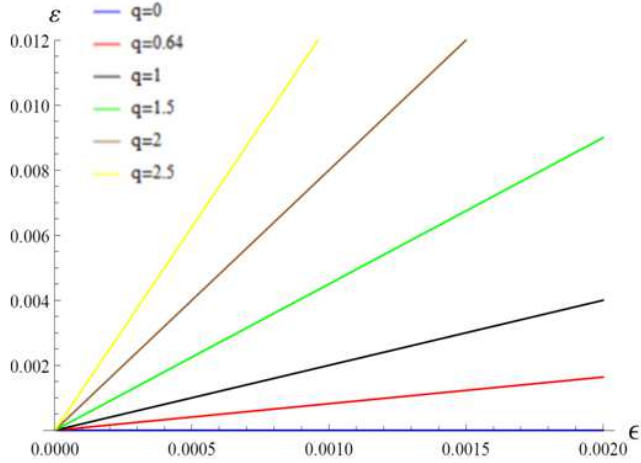,width=0.53\linewidth}\caption{\small{Plot of
energy $\varepsilon$ with thickness of the shell $\epsilon$(km).
Fixing $H=0.00001$, $D=1$.}}\label{b2}
\end{figure}\end{center}

The energy content within the shell is given as
\begin{eqnarray*}
\varepsilon&=\int^{D+\epsilon}_D 4 \pi \rho r^2dr
\end{eqnarray*}
which is found after using the corresponding values from the
equation of motion as follows
\begin{eqnarray}\label{38}
\varepsilon=\frac{4\pi
H}{7}\left[(D+\epsilon)^7-D^7\right]+2q^2\epsilon
\end{eqnarray}
The graphical representation about the role of energy and thickness
of the intermediate shell is being explored from the above equation
and is shown in Fig.\eqref{b2}. The graph \eqref{b2} shows the
linear relationship between energy and thickness of the shell, while
energy of the system tends to increase by increasing the
corresponding charge values.

\subsection{Entropy of the Charged Gravastars}

\begin{center}\begin{figure}\centering
\epsfig{file= 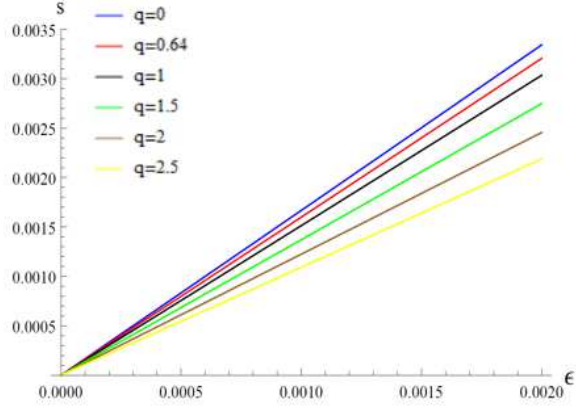,width=0.49\linewidth}\caption{\small{Plot of
entropy $S$ with thickness of the shell $\epsilon$(km). Fixing
$\chi=1$, $C=0.00006$, $H=0.00001$, $D=1$, $\alpha=1$.}}\label{b3}
\end{figure}\end{center}

Entropy is the disorderness within the body of a gravastar. It is found in the literature that
the entropy density of the interior region of the charged gravastar is zero. The entropy relation for the shell can
be calculated through the following formula
\begin{eqnarray}\label{39}
S=\int^{D+\epsilon}_D 4\pi r^2S(r)\sqrt{e^\lambda}dr,
\end{eqnarray}
where
\begin{eqnarray}\label{40}
S(r)=\frac{\alpha^2K_B^2 T(r)}{4\pi
h^2}=\alpha\left(\frac{K_B}{h}\right)\sqrt{\frac{P}{2\pi}},
\end{eqnarray}
describes entropy density corresponding to a specific temperature $T(r)$. In the above expression,
$\alpha$ is a constant term that has no any dimension. It is noteworthy that we are using geometrical ($G=C=1$) as well as Planck units ($K_B=\hbar=1$) in our computation, therefore $S(r)$ becomes
\begin{eqnarray}\label{41}
S(r)=\alpha\sqrt{\frac{P}{2\pi}}.
\end{eqnarray}
Then, Eq.\eqref{39} turns out to be
\begin{eqnarray}\label{42}
S=(8\pi
H)^\frac{1}{2}\alpha\int^{D+\epsilon}_D\frac{r^4}{\sqrt{\frac{q^2}{r^2}
+\frac{2\chi}{r^2}(\frac{1}{4\pi}+1)q^2+2lnr+C}}dr.
\end{eqnarray}
The above equation contains the contribution of charge as well as corrections from $f(R,T)$ gravity. The analytical solutions of the above is not possible. After using numerical method, we have drawn graphs to
examine the behavior of an electric charge which are given in Fig.\eqref{b3}.

\subsection{EoS Parameter}

At a particular radius $r=D$,the EoS can written as
\begin{eqnarray}\label{43}
\omega(D)=\frac{\nu}{\sigma}.
\end{eqnarray}
Substituting the values of $\nu$ and $\sigma$ in Eq.\eqref{43},
we get
\begin{eqnarray}\label{44}
\omega(D)=\frac{\left[\frac{\left(1-\frac{M}{D}\right)}{\sqrt{1-\frac{2M}{D}+\frac{q^2}{D^2}}}
-\frac{\left[1-\frac{8D^2\rho_0}{3}(2\pi+\chi)+
\frac{q^2}{2D^2}+\frac{\chi}{2D^2}\left(\frac{1}{4\pi}+1\right)
q^2\right]}{\sqrt{1-\frac{4D^2\rho_0}{3}(2\pi+\chi)+\frac{q^2}{D^2}+\frac{\chi}{D^2}
\left(\frac{1}{4\pi}+1\right)
q^2}}\right]}{-2\left[\sqrt{1-\frac{2M}{D}+\frac{q^2}{D^2}}-\sqrt{1-\frac{4D^2\rho_0}{3}
(2\pi+\chi)+\frac{q^2}{D^2}+\frac{\chi}{D^2}\left(\frac{1}{4\pi}+1\right)
q^2}\right]}.
\end{eqnarray}
To get real solutions of the above equation, we have a tendency to use some approximations, i.e,
$\frac{2M}{D}<1$ and $\frac{4(2\pi+\chi)\rho_0D^2}{3}<1$. We use
binomial expansion for avoiding square root terms which is
responsible for producing increments in the sensitivity of equation and some
approximations, i.e, $\frac{M}{D}<1$,
$\frac{4(2\pi+\chi)\rho_0D^2}{3}<1$ and $\frac{q^2}{2D^2}\ll1$, then
we get
\begin{eqnarray}\label{45}
\omega(D)\approx\frac{-3D
q^2+3Mq^2+12(2\pi+\chi)\rho_oD^5-2(2\pi+\chi)q^2\rho_oD^2[5
+6(\frac{1}{4\pi}+1)\chi]}{2D[-4(2\pi+\chi)\rho_oD^4+3\chi(\frac{1}{4\pi}+1)q^2+6M]},
\end{eqnarray}
which can be rewritten as
\begin{eqnarray}\label{46}
\omega(D)\approx\frac{\phi_1-\phi_2}{8D^5(2\pi+\chi)\rho_o\left[\phi_3-1\right]},
\end{eqnarray}
where
\begin{eqnarray}\nonumber
\phi_1&=&3Mq^2+12(2\pi+\chi)\rho_oD^5,\quad \phi_2=2(2\pi+\chi)q^2\rho_oD^2[5
+6(\frac{1}{4\pi}+1)\chi],\\\nonumber
\phi_3&=&\frac{3\chi(\frac{1}{4\pi}+1)q^2+6M}{4(2\pi+\chi)\rho_oD^4}.
\end{eqnarray}
The sign of EoS parameter is being controlled by the signs of numerator and denominator. The EoS parameter becomes positive, if $\phi_1>\phi_2$ along with $\phi_3>1$ or $\phi_1<\phi_2$ with $\phi_3<1$. However, if during evolution, stellar system satisfy the the constraints $\phi_1>\phi_2,~\phi_3<1$ or $\phi_1<\phi_2,~\phi_3>1$, then the $\omega$ will enter into a negative phase. For instance, the choice $\omega=-1$ incorporates the DE effects of
the cosmological constant $\Lambda$. This scenario could be helpful to understand the theoretical modeling of gravastars.

\section{Conclusion}

In this work, we have investigated the role of electromagnetic field
on an isotropic stellar model with extra degrees of freedom coming
from $f(R,T)$ gravity. Gravastar is the short form of Gravitationally vacuum
stars, that take up a new idea in the gravitational system. Such kind of
stellar model can be considered as an alternative to black
holes. Gravastar can be described through three different regions, first
is the interior region with radius $r$, second is the intermediate thin shell
with thickness $\epsilon$ and third is the exterior region with radius
$r+$. The evolution of fluid is dealt with by a specific EoS. We have worked
out a set of singularity-free solution of gravastar that represents different
features of the isotropic relativistic system. Some of the discussed properties
of our systems are described below.\\
(1) $Pressure-density~profile:$ The relationship between pressure
and density of the ultrarelativistic fluid within the intermediate
thin shell is shown in Fig.\ref{a1} against the radial coordinate
$r$. We can see the effect of electromagnetic charge on pressure and
density.\\
(2) $Proper~length~of~thin~shell:$ Figure \ref{b1} is plotted
between proper length of the shell and the thickness of
the shell. We can conclude from the graph that if charge within the gravastar is increasing then
length of thin shell is decreasing and if charge within the
gravastar is decreasing then length of thin shell is
increasing. The electromagnetic field and thickness of the shell are having an inverse relation.\\
(3) $Energy~content$: Energy within the shell and thickness of the
shell are directly proportional to each other. It can be seen from Fig.\ref{b2} that the increase of
charge energy would directly increase the thickness of the shell. Furthermore, the thickness can be enhanced by including huge amount of charge in gravastars. \\
(4) $Entropy:$ In order to see the role of entropy, thickness and electric charge, we have drawn a graph mentioned in Fig.\ref{b3}. This graph shows the linear
relationship between entropy and thickness of the shell. By studying
the effect of electromagnetic charge, we can conclude that if charge
in the gravastar is increasing then shell's entropy is decreasing and vice versa.\\
(5) $Equation~of~state:$ We use some approximation on binomial
expressions to required a real solution of $\omega(D)$. The constraints depends upon
the electric charge, $f(R,T)$ corrections, mass and radius of the metric.

We have an overall observation regarding the contribution of $f(R,T)$ gravity is that unlike GR the involvement of extra degrees of freedom coming from $\chi$ has made our analysis quite different in both mathematical and graphical point of view. On assigning zero value to this coupling constant would eventually provide the limiting case of GR.

\vspace{0.25cm}

{\bf Acknowledgments}

\vspace{0.25cm}

The works of Z. Yousaf and M.Z. Bhatti have been supported financially by National Research Project for Universities (NRPU), Higher Education Commission, Pakistan under research project No. 8754. In addition, the work of KB was supported in part by the JSPS KAKENHI Grant Number JP 25800136 and Competitive Research Funds for Fukushima University Faculty (18RI009).

\end{document}